\documentclass[prl,twocolumn,nowpacs,floatfix,amsmath,amssymb,superscriptaddress]{revtex4-1}
\usepackage{bibunits}
\usepackage{latexsym}
\usepackage{graphicx}
\usepackage{tabularx}
\usepackage{times}
\usepackage{color}
\usepackage{amsmath}
\usepackage{dcolumn}
\usepackage{latexsym,amsmath,amssymb,bm,euscript}

\begin{document}

\title{Revealing topological superconductivity in extended quantum spin Hall Josephson junctions}

\author{Shu-Ping Lee}
\affiliation{Department of Physics and Institute for Quantum Information and Matter, California Institute of Technology,
Pasadena, CA 91125, USA}
\author{Karen Michaeli}
\affiliation{Department of Condensed Matter Physics, Weizmann Institute of Science, Rehovot, 76100, Israel}
\author{Jason Alicea}
\affiliation{Department of Physics and Institute for Quantum Information and Matter, California Institute of Technology,
Pasadena, CA 91125, USA}
\author{Amir Yacoby}
\affiliation{Department of Physics, Harvard University, Cambridge, Massachusetts 02138 USA}

\begin{abstract}
{Quantum spin Hall-superconductor hybrids are promising sources of topological superconductivity and Majorana modes, particularly given recent progress on HgTe and InAs/GaSb.  We propose a new method of revealing topological superconductivity in extended quantum spin Hall Josephson junctions supporting `fractional Josephson currents'.  Specifically, we show that as one threads magnetic flux between the superconductors, the critical current traces an interference pattern featuring sharp fingerprints of topological superconductivity---even when noise spoils parity conservation.  }
\end{abstract}
\maketitle
\bibliographystyle{apsrev4-1}
\textbf{\emph{Introduction.}}\  `Spinless' one-dimensional (1D) topological superconductors \cite{1DwiresKitaev,BeenakkerReview,AliceaReview,FlensbergReview,StanescuReview} host various novel phenomena, most notably Majorana zero-modes that lead to non-Abelian statistics and, in turn, fault-tolerant quantum information applications \cite{TQCreview}.  Among numerous plausible realizations \cite{MajoranaQSHedge,1DwiresLutchyn,1DwiresOreg,CookFranz,Choy,Yazdani}, Fu and Kane's early proposal for nucleating 1D topological superconductivity at a quantum spin Hall (QSH)/superconductor interface remains a leading contender \cite{Disorder4}.  Experiments have, moreover, shown exciting recent progress, with QSH behavior \emph{and} good proximity effects conclusively demonstrated in both HgTe \cite{BHZ,Konig,EdgeTransportHgTe,Nowack1,HgTeProximity} and InAs/GaSb \cite{QuantumWellQSH,Knez,Nowack2,KnezProximity} quantum wells.

In light of these developments, the following question becomes paramount:
How can one compellingly reveal topological superconductivity in these QSH setups?  Most detection protocols to date focus on tunneling \cite{BeenakkerDetection1,BeenakkerDetection2,Crepin2014} and Josephson \cite{MajoranaQSHedge,LinderQSH,Badiane,JiangFractionalJosephson,dual_FJE_Meng,dual_FJE_Jiang,Houzet,BeenakkerWideJunction,Barbarino,Sun,Crepin,MeyerReview} anomalies.  The latter originate from the `fractional Josephson effect' \cite{1DwiresKitaev} wherein a phase twist $\delta\phi$ across a topological superconductor yields a supercurrent with $4\pi$ periodicity in $\delta\phi$---twice that of conventional junctions.  One can view the doubled periodicity as arising from a pair of hybridized Majorana modes at the junction, which form an unusual Andreev bound state that mediates supercurrent via single electron (rather than Cooper pair) tunneling.  In the simplest case this anomalous current takes the form $I_{4\pi} \propto (-1)^p \sin(\delta\phi/2)$, where the parity $p = 0,1$ denotes the Andreev bound state's occupation number.  Directly observing this spectacular effect is, however, nontrivial.  Parity switching processes---which send $p\rightarrow 1-p$ and can arise, \emph{e.g.}, from quasiparticle poisoning---restore $2\pi$ periodicity to the current unless measurements occur on a time scale short compared to the typical parity-flip time.  (Long-time-scale measurements may still reveal subtler signatures of topological superconductivity \cite{Badiane,Houzet,BeenakkerWideJunction,MeyerReview}, for instance through noise.)

\begin{figure}
\includegraphics[width = \columnwidth]{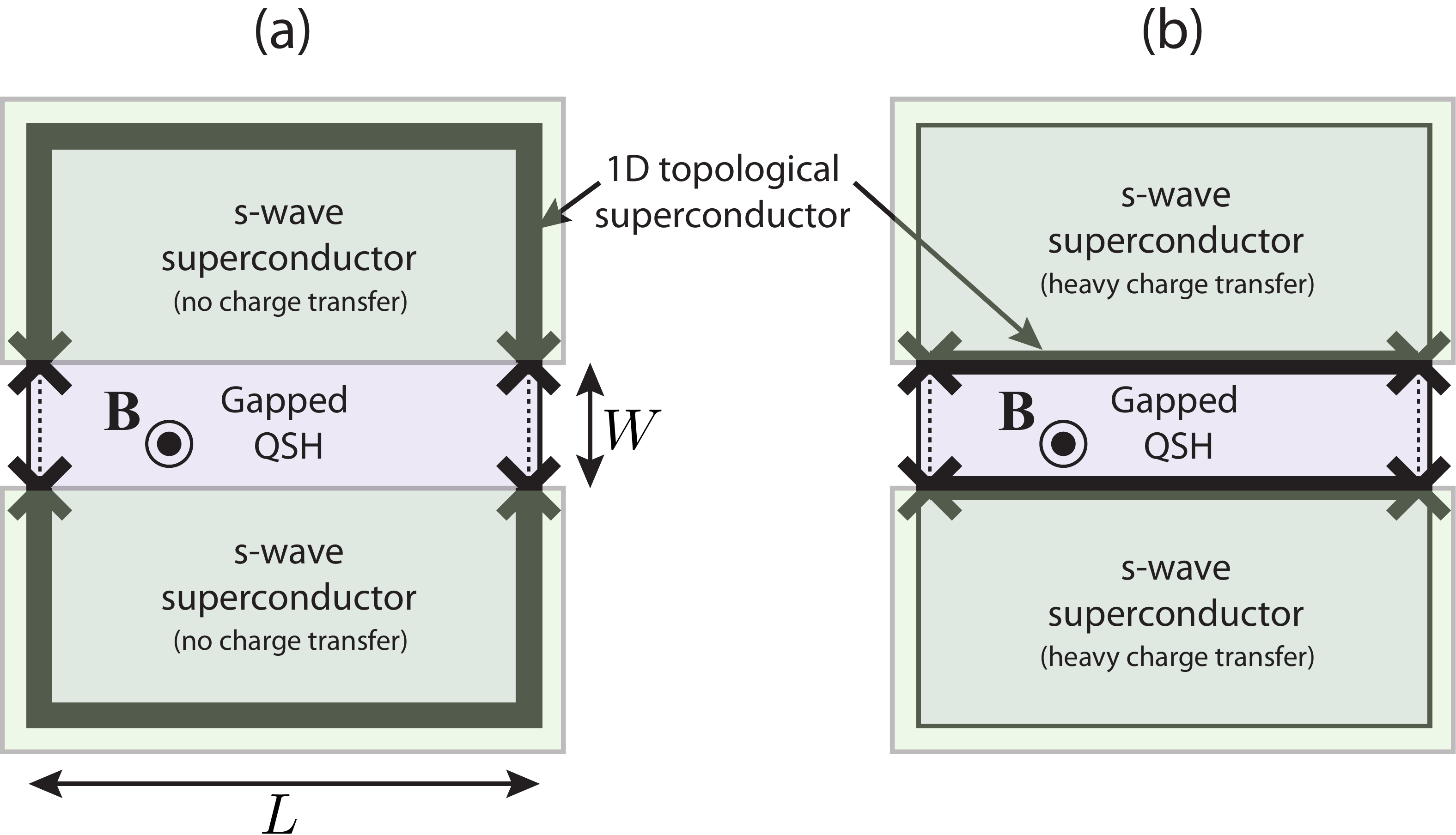}
\caption{(color online). Extended QSH Josephson junctions that host 1D topological superconductivity.  Topological superconductors reside either (a) at the outer boundary or (b) across the barrier depending on whether the superconductors dope the contacted QSH regions.  }
\label{Setups}
\end{figure}

Inspired by recent experiments by Hart \emph{et al}.\ \cite{HgTeProximity}, we study transport in an extended Josephson junction bridged by a QSH insulator; see Fig.\ \ref{Setups}.  This setup is expected to host two 1D topological superconductors that produce `parallel' fractional Josephson currents at the junction ends.  One virtue of such extended junctions is that the critical current $I_c(\Phi)$, measured as a function of magnetic flux $\Phi$ passing between the superconductors, displays an interference pattern that can reveal detailed information about the nature of current flow.
Here we ask whether such interference measurements can provide fingerprints of 1D topological superconductivity.

Our central result is that the fractional Josephson effect indeed imprints qualitative signatures of topological superconductivity on the junction's interference pattern and the corresponding critical current, \emph{even when parity switching processes are abundant}.  If parity relaxes to minimize the energy, the critical current remains finite at any magnetic flux contrary to conventional symmetric junctions.  Still more striking signatures appear if parity instead flips randomly on suitably long time scales---multiple critical currents are visible in the current-voltage traces, and the lower critical current \emph{vanishes} at zero flux provided the fractional Josephson currents dominate.  These results highlight relatively simple dc measurements that can reveal 1D topological superconductivity in QSH junctions and related platforms.

\textbf{\emph{Extended Josephson junction model.}}\ Following Ref.~\cite{HgTeProximity} we consider two $s$-wave superconductors deposited on a QSH material to create an extended Josephson junction of barrier width $W$ and length $L$ (see Fig.\ \ref{Setups}).  Suppose first that the QSH system's chemical potential resides everywhere in the bulk gap, and that the superconductors merely induce pairing via the proximity effect.  The edge states along the perimeter then form 1D topological superconductors \cite{MajoranaQSHedge} that hybridize at the junction as Fig.\ \ref{Setups}(a) illustrates. Effectively, the bulk behaves as an SIS junction, while the edges form two parallel SNS junctions that mediate the majority of the current.  This picture is supported by the interference pattern observed in HgTe junctions similar to those examined here\ \cite{HgTeProximity}.  Due to work-function mismatch, however, we expect that in practice the superconductors additionally transfer charge and shift the local Fermi level in the contacted QSH regions well into the bulk bands (though the barrier can still remain depleted).  In this scenario one can always change the sign of the mass for the heavily doped regions without closing a gap.  The outer regions then admit a trivial band structure---hence edge states occur \emph{only} at the boundary of the smaller QSH insulator comprising the barrier.  As shown in Fig.\ \ref{Setups}(b) these edge modes form 1D topological superconductors due to proximity with the adjacent superconductors; their hybridization yields the same physics as in Fig.\ \ref{Setups}(a).

For simplicity we assume negligible edge-state penetration depth and $W \ll \xi$ and $L \gg \xi$ throughout, with $\xi$ the coherence length of the 1D topological superconductors.  In this limit the left/right junction ends each support a single Andreev bound state with energy $(-1)^{p_{L/R}}\Delta \cos(\delta\phi_{L/R}/2)$.  Here $\Delta$ is the induced pairing energy while $p_{L/R}$ and $\delta\phi_{L/R}$ respectively denote the parity and phase difference at the left/right sides.  Generally, $\delta\phi_{L/R}$ follow from the phase difference $\phi$ between the two superconductors and the number of flux quanta $f = \Phi/(h/2e)$ threading the barrier---\emph{i.e.}, $\delta\phi_L = \phi$ and $\delta\phi_R = \phi + 2\pi f$.  Defining a vector ${\bf p} = (p_L,p_R)$, the bound states together contribute an energy
\begin{equation}\label{eq:Energy}
  E_{\bf p}(\phi,f)=\Delta[(-1)^{p_L}\cos(\phi/2)+(-1)^{p_R}\cos(\phi/2+\pi f)]
\end{equation}
and a Josephson current $I_{\bf p}(\phi,f) = \frac{e}{\hbar}\partial_\phi E_{\bf p}(\phi,f)$. Note that the bound-state energies merge with the continuum at isolated values of $\delta\phi_{L/R}$; thus, quasiparticles above the gap constitute one important parity-switching source.  One can, however, mitigate this particular switching mechanism by energetically isolating the bound states via in-plane magnetic fields~\cite{MajoranaQSHedge}, or with interactions in wider junctions \cite{ZhangKane}.

We consider a current-biased junction and extract the $I-V$ characteristics using an over-damped RCSJ model~\cite{Tinkham}.  The total injected current $I$ derives from two parallel channels: the Josephson current and resistive sources such as normal quasiparticles characterized by a resistance $R$.  The former---which we temporarily assume consists only of $I_{\bf p}$---shunts the resistive component $I_{N}=V/R = \hbar\dot\phi/(2eR)$ provided the junction does not generate voltage.  \emph{Between two parity-switching events} the phase $\phi$ thus evolves according to
\begin{equation}
  I=I_{\bf p}(\phi,f)+\frac{\hbar}{2eR}\dot\phi+\zeta(t),
  \label{eq:phiEOM}
\end{equation}
where the last term reflects a thermal noise current satisfying $\langle\zeta(t)\zeta(t')\rangle=2T/R\delta(t-t')$ ($T$ denotes the junction temperature; throughout we assume $T \ll \Delta$).  Equation \eqref{eq:phiEOM} describes a strongly damped particle with coordinate $\phi$ in a `tilted washboard' potential $U_{\bf p}(\phi,f) = E_{\bf p}(\phi,f)-\hbar I\phi/e$.  For sufficiently small $I$ the potential favors pinning the particle to one of its minima.  Random thermal noise allows the particle to escape over the barrier \cite{HalperinPhaseSlips}, whereupon the frictional force $\hbar\dot{\phi}/2eR$ causes relaxation to a new minimum on a time scale proportional to $\tau_R \equiv \hbar^2/(4e^2 R \Delta)$.  No minima exist above a parity-dependent critical current; the particle then rolls unimpeded down the potential, generating a substantial voltage.

Parity-switching events transfer the particle between different tilted washboard potentials ($U_{\bf p} \rightarrow U_{\bf p'}$) and thus provide an additional route for the phase $\phi$ to diffuse even at zero temperature.  Our goal now is to quantify the effects of parity switching on transport in various interesting regimes.

\textbf{\emph{Fokker-Planck analysis.}}  To this end let $\mathcal{P}_{\bf p}(\phi,t)$ be the distribution function that describes the probability of finding the system with parities ${\bf p}$ and phase $\phi$ at time $t$.  This function obeys a generalized Fokker-Planck equation:
\begin{align}\label{eq:FokkerPlanck}
  \partial_t{\mathcal{P}}_{\bf p}&= \frac{1}{\tau_R \Delta} \partial_\phi\left[\partial_\phi U_{\mathbf{p}}/2+T\partial_\phi\right]\mathcal{P}_{\bf p}
  \\ \nonumber
  &+\sum_{{\bf p}'}\left[W_{\mathbf{p}'\rightarrow\mathbf{p}}\mathcal{P}_{\mathbf{p}'}-W_{\mathbf{p}\rightarrow\mathbf{p}'}\mathcal{P}_{\mathbf{p}}\right],
\end{align}
where the first line describes thermal phase diffusion along the tilted washboard potential $U_{\bf p}$ with fixed parity ~\cite{HalperinPhaseSlips,Risken} while the second incorporates parity switching with rates $W_{{\bf p}\rightarrow {\bf p}'}$. Equation \eqref{eq:FokkerPlanck} implicitly assumes that parity-flip processes do not involve an instantaneous change in the phase $\phi$; this holds provided the time scale for such events is the shortest in the problem.  We further postulate a phenomenological parity-switching mechanism whereby a particle bath connected to the junction allows electrons to tunnel between the bound states and the continuum of bulk excitations, localized states, and other particle sources.  We model the corresponding transition rate from parity configuration $\mathbf{p}$ to $\mathbf{p}'$ by
\begin{eqnarray}\label{eq:TransitionRate}
  W_{\mathbf{p}\rightarrow\mathbf{p}'}(\phi,f)&=&\frac{n[(U_{\mathbf{p}'}(\phi,f)-U_{\mathbf{p}}(\phi,f))/T_b]}{\tau}
  \nonumber \\
  &\times& \left[\delta_{p_R,p_R'}\delta_{p_L,1-p_L'}+\delta_{p_R,1-p_R'}\delta_{p_L,p_L'}\right],
\end{eqnarray}
with $1/\tau$ the typical parity-switching rate, $n[x]=(e^{x}+1)^{-1}$ the Fermi distribution function, and $T_b$ the bath temperature (which can differ from the junction temperature $T$).  We only consider independent parity flips at the two junction sides---hence the Kronecker delta's in Eq.~\eqref{eq:TransitionRate}.  The transition rate follows from Fermi's golden rule (for details see Supplementary Material) where $1/\tau$ is the rate in which electrons transfer between the particle sources and the junction, and $T_b$ corresponds to the window of available energies carried by them.  The limit $T_b\sim{T}\ll\Delta$, for instance, describes hopping between the junction and localized states in the bulk \cite{MajoranaQSHedge}. In contrast, quasiparticles in the superconductor that can enter with a large energy range correspond to the limit $T_b\rightarrow\infty$. The latter includes the enhanced quasiparticle poisoning occurring when the bound states merge with the continuum.

The junction's $dc$ voltage $V$ is determined by stationary solutions of Eq.\ \eqref{eq:FokkerPlanck}.  More precisely, the Josephson relation along with Eq.\ (\ref{eq:phiEOM}) yield
\begin{eqnarray}\label{eq:AverageV}
  V &=& \frac{\hbar}{2e}\langle \dot\phi\rangle = \frac{\hbar}{2e}\sum_{\mathbf{p}}\int_0^{4\pi}d\phi \dot\phi \mathcal{P}_{\mathbf{p}}(\phi)
  \nonumber \\
  &=& R\sum_{\mathbf{p}}\int_0^{4\pi}d\phi[I-I_{\bf p}(\phi,f)]\mathcal{P}_{\mathbf{p}}(\phi).
\end{eqnarray}
Determining the $I-V$ characteristics thus reduces to solving Eq.~\eqref{eq:FokkerPlanck} for the steady-state distribution function $\mathcal{P}_{\bf p}(\phi)$, which is readily achieved numerically by descretizing $\phi$.  Below we briefly discuss the solution with conserved parity ($1/\tau = 0$) and then address the more realistic case where parity switching occurs.

\begin{figure}
\includegraphics[width = \columnwidth]{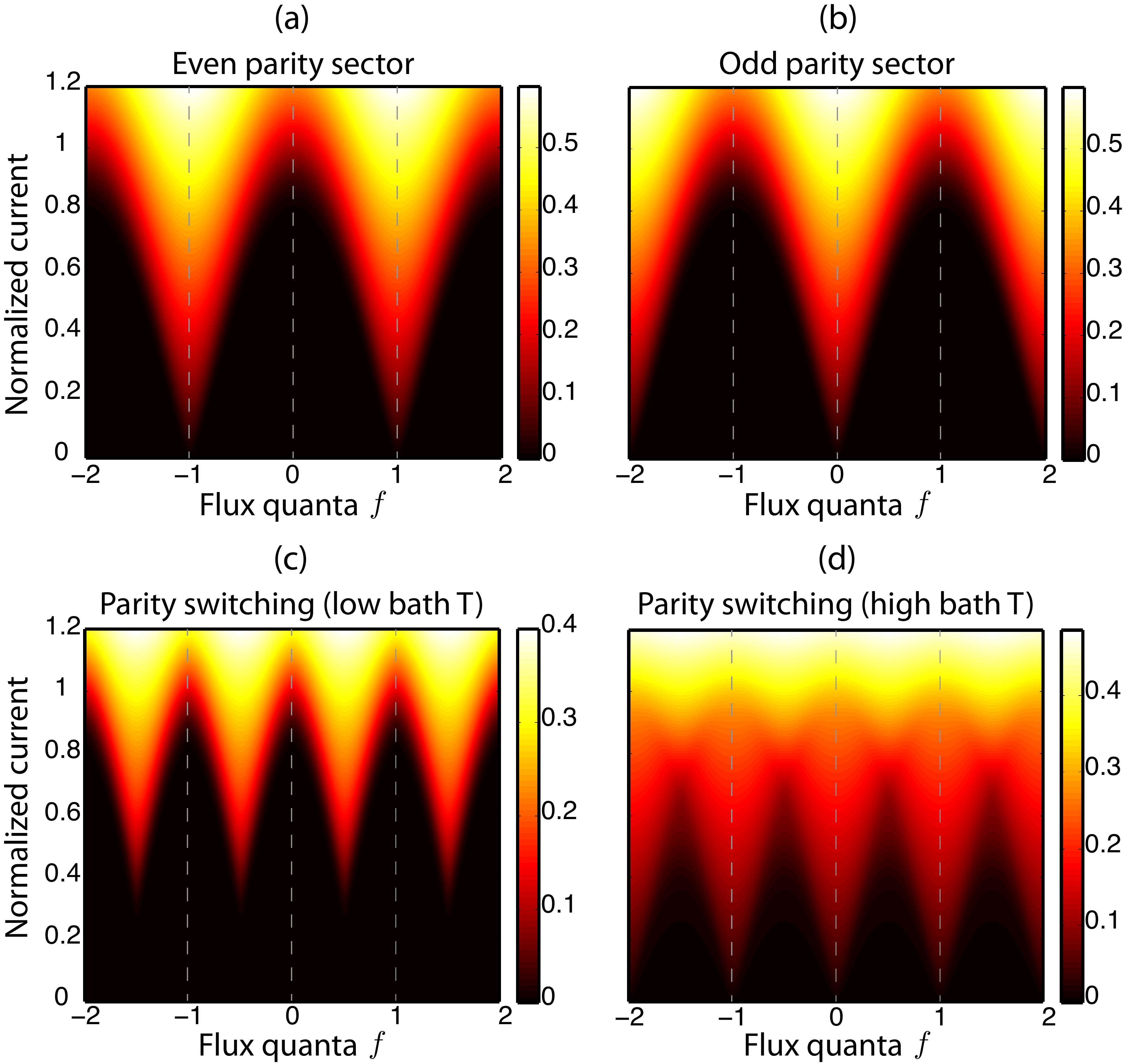}
\caption{(color online).  Interference patterns in (a,b) the parity-conserving limit and (c,d) with parity switching at low ($T_b = 0.02 \Delta$) and high ($T_b = 100\Delta$) bath temperature.  The color scale indicates voltage in units of $2eR\Delta/\hbar$ while current is normalized by $e\Delta/\hbar$.  Data correspond to (a,b) $T = 0.05\Delta$, (c) $T = 0.02\Delta$, $\tau = 50 \tau_R$, and (d) $T = 0.02\Delta$, $\tau = 5\tau_R$.  }
\label{Voltage_fig}
\end{figure}

When the parities ${\bf p}$ are conserved the generalized Fokker-Planck equation admits four steady-state solutions---one for each parity sector.  The solutions coincide with the known Ambegaokar-Halperin expressions \cite{HalperinPhaseSlips} evaluated with an unconventional current-phase relation $I_{\bf p}(\phi,f)$.  At $T = 0$ the voltage follows as~\cite{HalperinPhaseSlips,Tinkham}
\begin{align}\label{eq:IV}
  V = \Theta(I-I_{\mathbf{p},c})R\sqrt{I^2-I_{{\bf p},c}^2},
\end{align}
where $\Theta(x)$ is the Heaviside step function and the critical currents are $I_{{\bf p},c}={e\Delta}|\cos(\pi f/2)|/{\hbar}$ for $p_R=p_L$ and $I_{{\bf p},c}={e\Delta}|\sin(\pi f/2)|/{\hbar}$ for $p_R{\neq}p_L$.  Thermally induced $4\pi$ phase slips at fixed ${\bf p}$ produce a finite voltage at $T > 0$ even for $I < I_{{\bf p},c}$.  Figures \ref{Voltage_fig}(a) and (b) respectively illustrate the low-temperature interference patterns in the even- and odd-parity sectors (the color scale represents the voltage $V$).  Both cases exhibit an anomalous two-flux-quanta periodicity---a striking yet fragile fingerprint of topological superconductivity.  Indeed this property is spoiled by any finite switching rate $1/\tau \neq 0$, which in our setup will \emph{always} arise due to mixing with continuum quasiparticles and other noise sources.   Fortunately, as we now describe other signatures of topological superconductivity nevertheless persist.

For $1/\tau \neq 0$ Eq.\ \eqref{eq:FokkerPlanck} admits only one stationary solution due to parity flip processes.  Consider first $T_b \ll \Delta$ where the transition rates in Eq.\ \eqref{eq:TransitionRate} depend strongly on the relative energies in different parity sectors.  The behavior then resembles that of a thermalized junction: to a  good approximation parities switch only at energy crossings and adjust so that the system follows a washboard potential $U(\phi,f)=\min_{\mathbf{p}}U_{\mathbf{p}}(\phi,f)$ corresponding to a minimum energy.  The $T\rightarrow 0$ and $T_b \rightarrow 0$ critical current---\emph{i.e.}, the maximal $I$ for which $\partial_\phi U_{\mathbf{p}}(\phi,f)=0$ admits a solution---follows as $I_c=e\Delta/\hbar\max\{\cos^2(\pi f/2),\sin^2(\pi f/2)\}$.  Figure \ref{Voltage_fig}(c) displays the numerically computed interference pattern at small but finite $T$ and $T_b$ (which includes thermal phase slips that smear the critical current, as in conventional junctions).  The critical current clearly remains finite for all fluxes and, roughly, follows the larger of the critical currents present in the parity-conserving cases shown in Figs.\ \ref{Voltage_fig}(a) and (b).  Here the absence of nodes is a remnant of the unconventional current-phase relation rooted in topological superconductivity.  Other node-lifting sources also of course exist but can be distinguished from this mechanism as discussed below.

\begin{figure}
\includegraphics[width = \columnwidth]{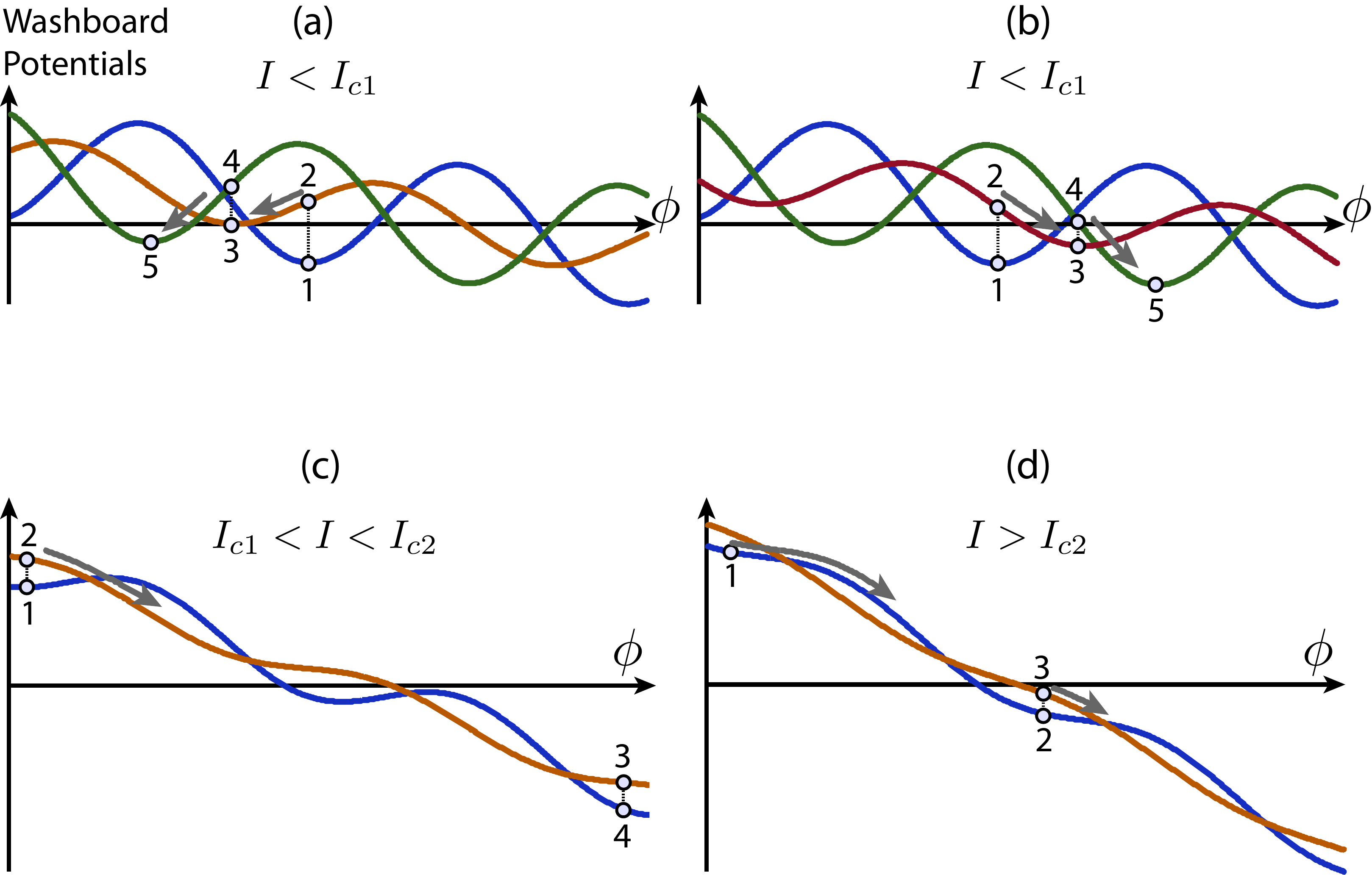}
\caption{(color online). Washboard potentials for select parity sectors in the three high-bath-temperature current regimes.  For low currents $I< I_{c1}$ consecutive parity flips can mediate $\pm 2\pi$ phase slips as in (a) and (b).  In (c) and (d) a steady phase drift always occurs.}
\label{washboard_fig}
\end{figure}

Finally, we analyze the most interesting limit---$T_b \gg \Delta$ where the parities fluctuate randomly, independent of the initial and final energies, on a time scale $\tau$.  Here there are three distinct current regimes separated by the critical currents $I_{c1}=\min_{\bf p}I_{{\bf p},c}$ and $I_{c2}=\max_{\bf p}I_{{\bf p},c}$.  For $I<I_{c1}$ local minima exist in the washboard potentials $U_{\bf p}$ for all four parity sectors. Nevertheless, even at $T = 0$---where thermal diffusion is absent---the phase $\phi$ can still transform between minima of $U_{\bf p}$ via parity-switching events; see Figs.\ \ref{washboard_fig}(a) and (b).  The voltage resulting from such processes depends on the ratio of $\tau$ to the typical time $\tau_{\rm rel}$ required for $\phi$ to relax to a washboard-potential minimum following a parity flip:
\begin{align}\label{eq:t_rel}
  \tau_{\rm rel}\sim\max_{\bf p}\frac{\hbar}{eR\sqrt{I_{{\bf p},c}^2-I^2}}.
\end{align}
(A similar time scale emerges in the $ac$ fractional Josephson effect~\cite{MeyerReview}.)

For $\tau \gg \tau_{\rm rel}$ the phase $\phi$ has sufficient time to reach the nearest minimum of the new potential before parity switches again.  After two consecutive parity flips $\phi$ either returns to its initial value or, as Figs.\ \ref{washboard_fig}(a) and (b) illustrate, shifts by $\pm 2\pi$.  The $2\pi$ and $-2\pi$ phase changes occur with essentially equal probability when $T_b \gg \Delta$, and moreover contribute equal but opposite voltages.  Hence these processes cancel one another in the $dc$ limit.  In other words, parity switching events generate telegraph noise in the voltage with equal probability of positive and negative signals that time-average to zero.   As the current approaches $I_{c1}$, the relaxation time $\tau_{\rm rel}$ grows and eventually exceeds the parity-flip time $\tau$.  Consecutive switching events then occur before the phase relaxes to the potential minima; the result is a net diffusion of $\phi$ down the washboard potentials, producing a finite voltage.  This argument implies that in the limit $\tau\lesssim\hbar/(eR I_{c1}) \sim\tau_R$ any current generates a non-zero voltage---\emph{i.e.}, the critical current vanishes.

With currents between $I_{c1}$ and $I_{c2}$ only two of the washboard potentials exhibit stable minima.  Because of the high bath temperature, the phase $\phi$ can escape from one of these minima via a parity-switching event into a potential without any minima, producing a steady drift of $\phi$.  The drift ceases only when a subsequent parity flip re-traps the phase; see Fig.\ \ref{washboard_fig}(c) for an illustration.  Assuming $\tau \gg \tau_{\rm rel}$, the phase drift generates a finite $dc$ voltage $V\approx \mathcal{F}_{\rm drift} R\sqrt{I^2-I_{c1}^2}$ with $\mathcal{F}_{\rm drift}$ the fraction of time spent in potentials without minima ($\mathcal{F}_{\rm drift} \approx 1/2$ when $T_b \gg \Delta$).   For currents close to $I_{c2}$ the phase relaxation time $\tau_{\rm rel}$ exceeds $\tau$; the phase can then essentially never reach a minimum due to frequent parity flips.  An additional voltage contribution thus appears, which smears the voltage as a function of current near $I_{c2}$---just as in the region near $I_{c1}$ discussed earlier.

Above $I_{c2}$ none of the bands support minima, and the phase $\phi$ drifts continuously as in Fig.~\ref{washboard_fig}(d).  The instantaneous drift velocity and hence also the voltage are nonetheless parity dependent.  It follows that parity switching events, on average, produce a voltage $V \approx \mathcal{F}_{\rm drift}'R\sqrt{I^2-I_{c1}^2}+(1-\mathcal{F}_{\rm drift}')R\sqrt{I^2-I_{c2}^2}$.  Here $\mathcal{F}_{\rm drift}'$ and $1-\mathcal{F}_{\rm drift}'$ denote the fraction of time the phase spends in the potentials with critical currents $I_{c1}$  and $I_{c2}$, respectively.

\begin{figure}
\includegraphics[width = \columnwidth]{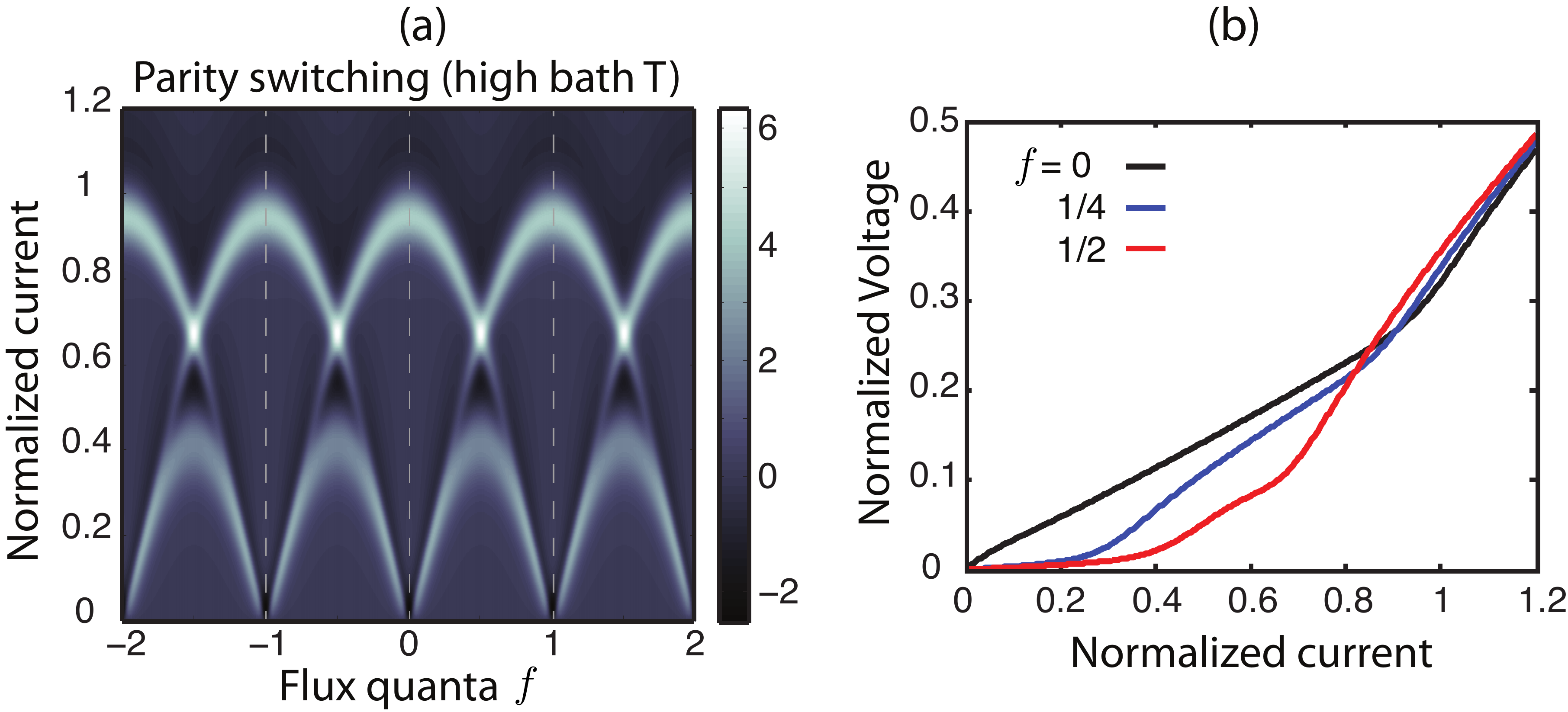}
\caption{(color online).  (a) Color plot of $d^2V/dI^2$ and (b) voltage-current line cuts corresponding to the high-bath-temperature data in Fig.\ \ref{Voltage_fig}(d). The two critical currents $I_{c1}$ and $I_{c2}$ are clearly visible in both plots.  Voltage and current are respectively expressed in units of $2eR\Delta/\hbar$ and $e\Delta/\hbar$.}
\label{TwoCriticalCurrents}
\end{figure}

We thus arrive at the following overall picture for the high-bath-temperature case.  When $\tau \gg \tau_R$ the $dc$ voltage remains negligible as long as $I<I_{c1}=\min_{\bf p}I_{{\bf p},c}$.  That is, contrary to the limit $T_b \ll \Delta$ the (lower) critical current as a function of flux follows the \emph{minimum} of the critical currents associated with the four parity sectors.  Furthermore, the critical current vanishes at zero flux and is maximal at one-half flux quantum---precisely as in a $\pi$-junction [see Fig.\ \ref{Voltage_fig}(d)].  For $I> I_{c1}$ the voltage is far from featureless---a second critical current $I_{c2}=\max_{\bf p}I_{{\bf p},c}$ also appears, reflecting the multiple parity sectors.  This feature becomes prominent upon examining $d^2V/dI^2$ [Fig.\ \ref{TwoCriticalCurrents}(a)] as well as specific voltage-current line cuts [Fig.\ \ref{TwoCriticalCurrents}(b)].  Thus long parity-flip times $\tau$ allow one to image the critical currents in all parity sectors.  Rapid parity flipping with $\tau \lesssim \tau_R$, however, renders the junction resistive at any flux and yields identically zero critical current.

\textbf{\emph{Discussion.}}\ Our study of extended QSH Josephson junctions reveals that parity switching processes, although destructive to the critical current's anomalous periodicity, generate new fingerprints of the underlying topological superconductors expected to form.  Surprisingly, stronger poisoning actually \emph{enhances} the signatures in the critical current (as long as parity fluctuates on sufficiently long time scales).  We expect the results to apply quite generally---even when the actual switching mechanism differs from our model.  For instance, if the bound-state energies approach the continuum states near $\Delta$ then bulk quasiparticles can easily mediate parity flips \cite{MeyerReview}.  We verified numerically that qualitatively similar behavior to the high-$T_b$ limit arises when switching occurs predominantly at energies near $\Delta$.

While our analysis has so far included only $4\pi$-periodic current contributions, it is important to note that conventional $2\pi$-periodic components $\propto \sin\delta\phi_{R/L}$ generically flow in parallel (though their magnitudes may be small).  The Supplementary Material addresses the consequences of such terms.  With low bath temperatures their effects are decidedly minor---the lifted nodes in Fig.\ \ref{Voltage_fig}(c) survive even for quite large conventional currents.  More significant effects occur at high bath temperature. There, the new terms lead to deviations from the $\pi$-junction behavior mimicked in Fig.\ \ref{Voltage_fig}(d).  The resulting interference pattern nevertheless \emph{still} remains anomalous.  Most importantly multiple critical currents remain visible in the current-voltage relation.  The critical current, as with low bath temperatures, also remains finite for any magnetic field.

The absence of nodes in the critical current at half-integer flux quanta thus survives quite generally from the interplay between fractional Josephson physics and parity switching (we include a tentative comparison with experiment regarding this feature in the Supplementary Material; see also Ref.\ \cite{PotterFu} for a somewhat related mechanism).  To provide a compelling indicator of topological superconductivity, however, the ability to experimentally distinguish from other node-lifting mechanisms such as current asymmetry is essential.  This may be achieved by introducing a strong \emph{in-plane} magnetic field, which can force the 1D topological superconductors at the junction into a trivial phase \cite{MajoranaQSHedge}.  Therefore, observing the controlled destruction and revival of nodes as one varied the in-plane field strength would likely rule out alternative mechanisms and provide strong evidence for topological superconductivity.

\textbf{\emph{Acknowledgements.}}\  We are indebted to D.\ Clarke, J.\ Meyer, J.\ Sau, A.\ Stern, D.\ van Harlingen, and especially B.\ Halperin, S.\ Hart, and H.\ Ren for enlightening discussions.
We also acknowledge funding from the NSF through grants DMR-1341822 (S.-P.\ L.\
\& J.\ A.) and  DMR-1206016 (A.\ Y.); the Alfred P.~Sloan Foundation (J.~A.); a grant from Microsoft Corporation (A.Y.); and the Caltech Institute for Quantum
Information and Matter, an NSF Physics Frontiers Center with support of the
Gordon and Betty Moore Foundation.  A.\ Y.\ is also supported by the STC Center for Integrated Quantum Materials, NSF grant DMR-1231319.

\section{Supplementary Material}

\subsection{1. Derivation of the generalized Fokker-Planck equation and the parity switching rate}

The starting point for deriving the rate equation for the distribution function is the Master equation. The Master equation is based on the assumption that if one considers the density matrix, its non-diagonal elements can be neglected; only the diagonal ones corresponding  to the distribution function are kept. In our case such a treatment is justified due to the macroscopic nature of the system.  For the Josephson junctions discussed here, the distribution function depends on two variables: the phase $\phi$ and the parity vector $\mathbf{p}$. Assuming independent transitions in these variables, the Master equations reads
\begin{align}\label{eq:Master}
  \partial_t{\mathcal{P}}_{\bf p}(\phi)&=
  \sum_{\phi'}\left[w_{\phi'\rightarrow\phi}\mathcal{P}_{\mathbf{p}}(\phi')-w_{\phi\rightarrow\phi'}\mathcal{P}_{\mathbf{p}}(\phi)\right]
  \\\nonumber&+\sum_{{\bf p}'}\left[W_{\mathbf{p}'\rightarrow\mathbf{p}}\mathcal{P}_{\mathbf{p}'}(\phi)-W_{\mathbf{p}\rightarrow\mathbf{p}'}\mathcal{P}_{\mathbf{p}}(\phi)\right].
\end{align}
Here, $w_{\phi\rightarrow\phi'}$ and $W_{\mathbf{p}\rightarrow\mathbf{p}'}$ are the transition rates between two different phase and parity states, respectively.   For the continuous variable $\phi$ the over-damped RCSJ model yields
\begin{align}\label{eq:Phase}
\sum_{\phi'}\left[w_{\phi'\rightarrow\phi}\mathcal{P}_{\mathbf{p}}(\phi')-w_{\phi\rightarrow\phi'}\mathcal{P}_{\mathbf{p}}(\phi)\right]\\\nonumber
\rightarrow\frac{1}{\tau_R \Delta} \partial_\phi\left[\partial_\phi U_{\mathbf{p}}/2+T\partial_\phi\right]\mathcal{P}_{\bf p}.
\end{align}
Thus the first term in Eq.~\eqref{eq:Master} is familiar from the conventional Fokker-Planck equation \cite{HalperinPhaseSlips}.

Next we will concentrate on the derivation of the parity transition rates. For this purpose we assume that the bound states connect to a particle source through a term in the Hamiltonian of the form $H_{t}=t\sum_{i=L,R}(d_i^{\dag}f+f^{\dag}d_i)$, where $d_{i}$ ($f$) is the annihilation operator for an electron in bound state $i$ (particle source). The index $i=L/R$ indicates the left and right bound state, so that the corresponding parity is $p_{i}=d_{i}^{\dag}d_{i}$.  The rate at which electrons transform from the particle bath to the bound state or vice versa can be approximated using Fermi's golden rule.  For example, if $|p_L,p_R\rangle$ denotes the parity eigenstate for the junction then
\begin{eqnarray}\label{eq:Parity1}
  &&W_{(0,p_R)\rightarrow(1,p_R)} (\phi)= 2\pi{t}^2|\langle{1,p_R|d_L^{\dag}}|0,p_R\rangle|^2
  \nonumber \\
  &&\hspace{5mm}\times\int{d\omega}g(\omega)n(\omega)\delta(U_{(1,p_R)}-U_{(0,p_R)}-\omega).
\end{eqnarray}
The above equation describes absorption of an electron from the particle bath by the left bound state (a similar term can be written for the right bound state). Here $n(\omega)$ is the distribution function of the electron poisoning source and $g(\omega)$ is its density of states, with $\omega$ measured with respect to the chemical potential. Possible sources of electrons that can hop into the bound states include---among others---subgap localized states and localized electrons in the bulk of the quantum spin Hall (QSH) system. These different sources are uncorrelated and therefore can be modeled as an incoherent particle bath with a constant density of state, $g(\omega)\approx{g}_0$.  The transition rate in Eq.~\ \eqref{eq:Parity1} therefore becomes
\begin{eqnarray}\label{eq:Parity2}
W_{(0,p_R)\rightarrow(1,p_R)} &=& \frac{n[(U_{(1,p_R)}-U_{(0,p_R)})/T_b]}{\tau}.
\end{eqnarray}
with $1/\tau=2\pi{g_0}t^2$.  Similarly, the transition rate for a particle hopping from the bound state into the bath reads:
\begin{eqnarray}\label{eq:Parity3}
&&W_{(1,p_R)\rightarrow(0,p_R)}= 2\pi{t}^2|\langle0,p_R|d_L|1,p_R\rangle|^2\int{d\omega}g(\omega)
  \nonumber \\
  &&\times\left[1-n(\omega)\right]\delta(U_{(1,p_R)}-U_{(0,p_R)}-\omega)
  \nonumber \\
  &&=\frac{n[(U_{(0,p_R)}-U_{(1,p_R)})/T_b]}{\tau}.
\end{eqnarray}
Here we used the identity $1-n(\omega)=n(-\omega)$. An equivalent treatment for the right bound state recovers the transition rates $W_{{\bf p}\rightarrow {\bf p}'}$ quoted in the main text.

Additional processes can contribute to the parity switching. For example, an electron can join a particle occupying the bound state and form a Cooper pair that hops into one of the superconducting leads. This process changes the parity from $(1,0)$ to $(0,0)$ by an absorption of a particle. The effect of such an event is to modify $1/\tau$ without changing the exponential term.

\subsection{2. Influence of conventional supercurrent}

The analysis performed in the main text assumed that the $4\pi$-periodic fractional Josephson currents $I_{\bf p}(\phi,f)$ constituted the only supercurrent source in our extended QSH Josephson junction.  We now analyze the more realistic case where ordinary Josephson currents---arising from continuum modes and/or additional Andreev bound states---flow in parallel.  We will assume that these contributions, like the fractional Josephson currents, reside solely along the left and right junction ends in Fig. 1 of the main text. Thus the total current for parity sector ${\bf p}$ is taken to be
\begin{equation}
  \tilde I_{\bf p}(\phi,f) = I_{\bf p}(\phi,f) + I_{2\pi}(\phi,f),
\end{equation}
where
\begin{equation}
  I_{2\pi}(\phi,f) = \frac{e\Delta_{2\pi}}{\hbar}[\sin\phi + \sin(\phi + 2\pi f)]
  \label{ConventionalCurrent}
\end{equation}
is the $2\pi$-periodic component with an associated energy scale $\Delta_{2\pi}$.  In terms of the energy $E_{2\pi}(\phi,f) = - \Delta_{2\pi}[\cos\phi+\cos(\phi+2\pi f)]$ for states mediating the current in Eq.\ (\ref{ConventionalCurrent}), the tilted washboard potentials that appear in the generalized Fokker-Planck equation are correspondingly modified to
\begin{equation}
  \tilde U_{\bf p}(\phi,f) = E_{\bf p}(\phi,f) +E_{2\pi}(\phi,f) -\frac{\hbar I \phi}{e}.
  \label{ModifiedU}
\end{equation}
Note that the transition rates $W_{{\bf p}\rightarrow {\bf p}'}$ depend only on the unconventional bound-state energies $E_{\bf p}(\phi,f)$ since all other terms in the modified potentials are parity-independent.

Let us first revisit the low-bath-temperature limit $T_b \ll \Delta$.  Here the parities again adjust such that the phase essentially follows the washboard potential $\tilde U(\phi,f) = \min_{\bf p}\tilde U_{\bf p}(\phi,f)$ that minimizes the total energy.  In this regime non-zero $\Delta_{2\pi}$ renormalizes the ratio of the maximum and minimum critical currents in the interference pattern, but importantly preserves the lifted nodes that are indicative of topological superconductivity.  At $f = (2n+1)/2$ for integer $n$---where the node lifting arises---the conventional current indeed drops out entirely since $I_{2\pi}[\phi,f = (2n+1)/2] = 0$.

The interplay between these two types of Josephson currents is more interesting at high bath temperature $T_b \gg \Delta$, where the system spends roughly equal time in all four parity sectors.  Subsequent parity flips occur after a typical time scale $\tau$ which we will assume greatly exceeds $\tau_R$.  For concreteness we further assume that the energy scales for conventional and Josephson currents are comparable (other limits can be treated similarly).  In this regime finite $I_{2\pi}$ leads to even richer structure in the current-voltage characteristics than in the $I_{2\pi} = 0$ case.

\begin{figure}
\includegraphics[width = \columnwidth]{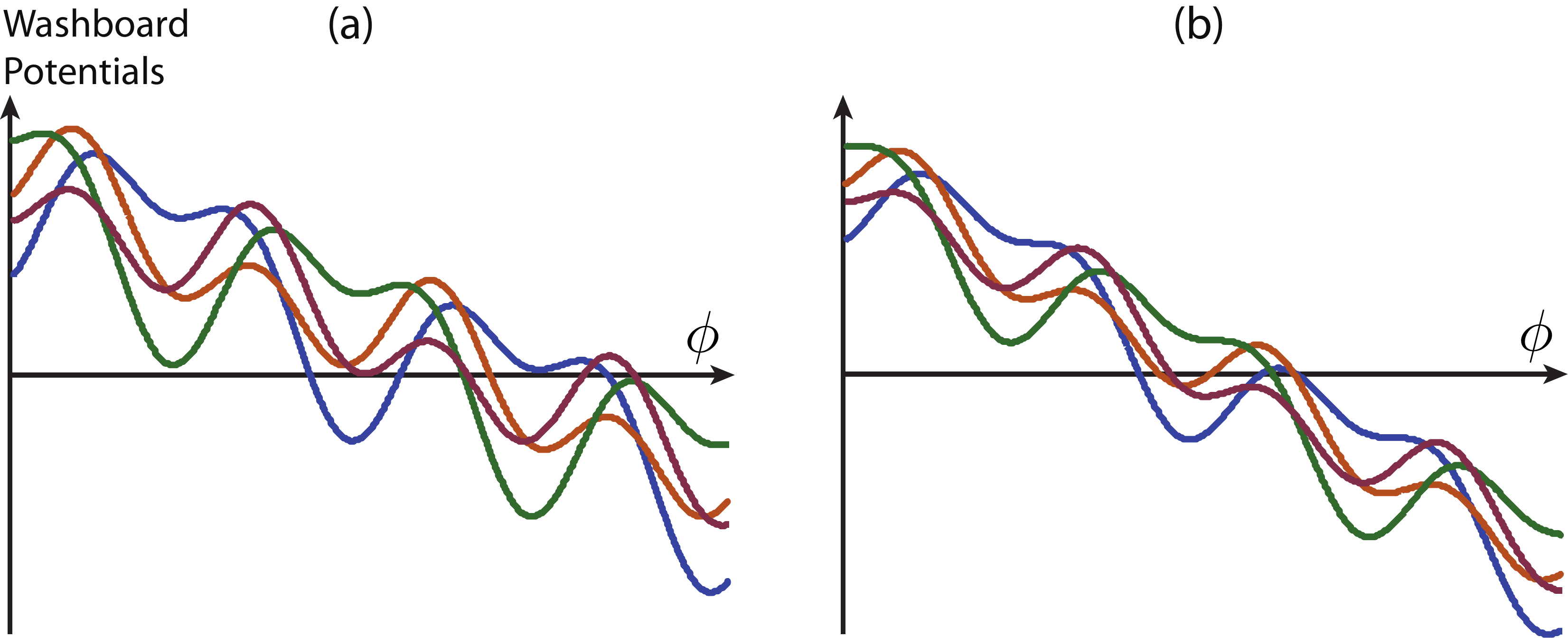}
\caption{(color online).  Tilted washboard potentials including a $2\pi$-periodic current component corresponding to $\Delta_{2\pi} = 0.8\Delta$ and $f = 1/4$.  Parts (a) and (b) correspond to different applied currents.  In (a) the current is sufficiently low that all potentials exhibit local minima roughly spaced by a phase difference of $2\pi$ due to the ordinary Josephson current.  In (b) the blue and green potentials possess half as many local minima due to a higher applied current.  Parity switching thus causes a slow drift of $\phi$ down the potentials so that a small $dc$ voltage develops.}
\label{washboards_supp}
\end{figure}

The additional structure can be anticipated upon examining the modified washboard potentials in Eq.\ \eqref{ModifiedU}.  Consider the lowest current regime, where for a given ${\bf p}$ the potential $\tilde U_{\bf p}(\phi,f)$ features denser minima spaced by a phase difference of roughly $2\pi$ (instead of $4\pi$ as is the case with only fractional Josephson currents).  As an example Fig.\ \ref{washboards_supp}(a) displays the four washboard potentials at $f = 1/4$, $\Delta_{2\pi} = 0.8\Delta$, and low current.  It is clear from the figure that parity flips no longer directly mediate $\pm 2\pi$ phase slips due to the additional minima; phase slips instead require thermal activation over a barrier.

At slightly larger currents the additional local minima generated by $I_{2\pi}$ disappear in some or all of the washboard potentials---see Fig.\ \ref{washboards_supp}(b) which corresponds to the same parameters as (a) except for a higher current.  Since all four washboard potentials still support local minima, the phase $\phi$ becomes trapped after each parity flip.  Parity switching does, nevertheless, mediate a gradual drift down the potentials and hence produces a (small) $dc$ voltage.  For instance, beginning from a minimum of the red curve, a parity flip into the blue potential can cause the phase to wind rightward.  Subsequent parity flips into the green curve will then cause a winding in the same direction.  `Upstream' drifting of the phase requires thermal activation and is thus suppressed compared to these processes.  The magnitude of the $dc$ voltage scales roughly as $V \sim \frac{eR \Delta}{\hbar}\frac{\tau_R}{\tau}$, where $\tau_R/\tau$ captures the fraction of time during which the phase drifts.

As the current further increases a situation familiar from the main text arises: Half of the potentials lose \emph{all} local minima leading to a much larger $dc$ voltage since the phase drifts unimpeded in certain parity sectors.  And finally in the highest current regime none of the potentials possess minima, and an even stronger voltage develops.

\begin{figure*}
\includegraphics[width = 15cm]{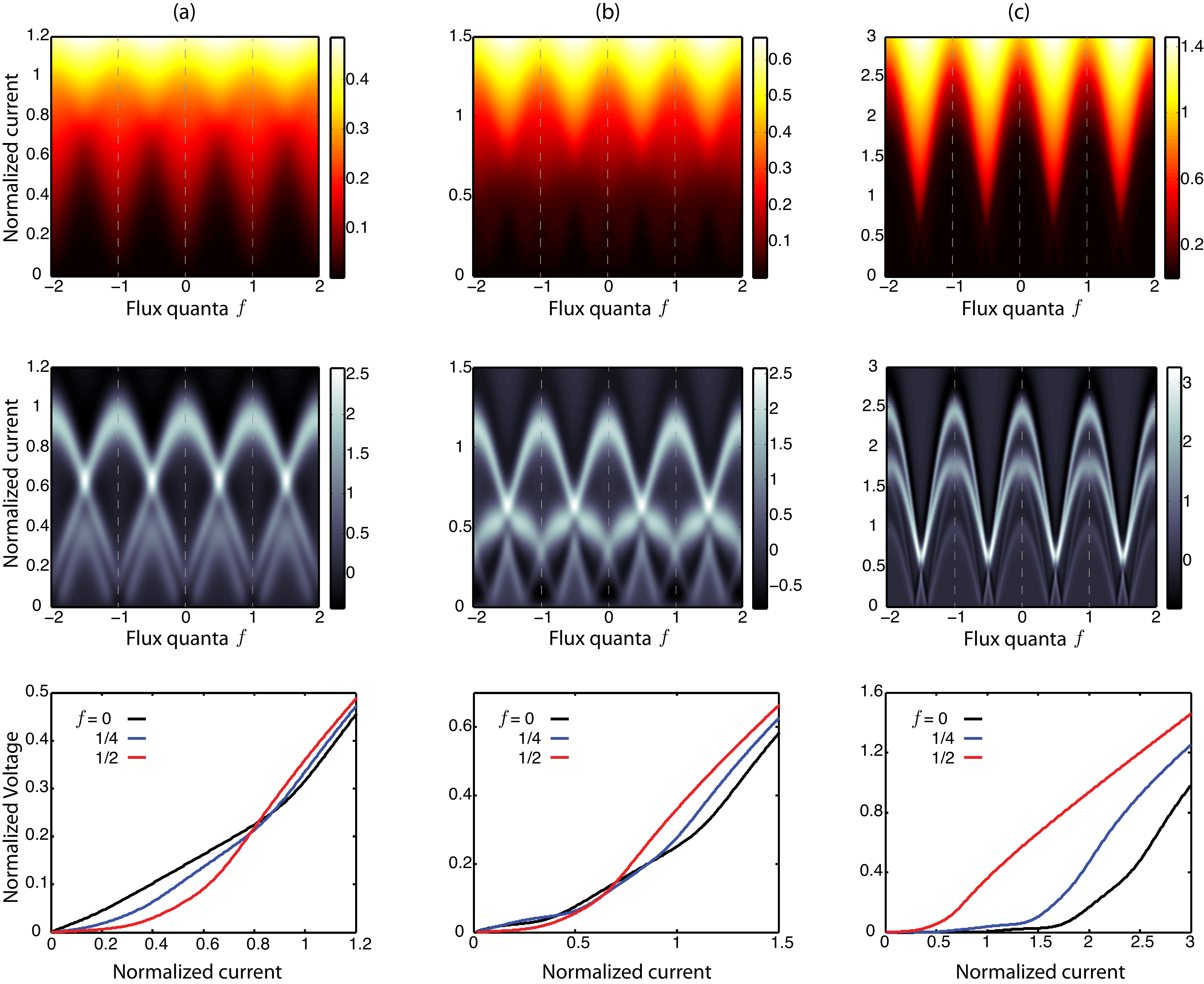}
\caption{(color online).  Current-voltage characteristics reflecting the interplay between fractional and conventional Josephson currents at high bath temperature $T_b = 100\Delta$ and with a long parity relaxation time $\tau = 50 \tau_R$.  The color scale in the first and second rows indicates voltage $V$ and its second derivative $d^2 V/dI^2$, respectively.  The last row illustrates voltage-current line cuts at specific flux values.  Going left to right the ordinary Josephson current increases, taking on  values corresponding to (a) $\Delta_{2\pi} = 0.1\Delta$, (b) $\Delta_{2\pi} = 0.25\Delta$, and (c) $\Delta_{2\pi} = \Delta$.  In all panels the junction temperature is $T = 0.05\Delta$, voltage is expressed in units of $2eR\Delta/\hbar$, and current is normalized by $e\Delta/\hbar$.  }
\label{ConventionalCurrents}
\end{figure*}

To summarize, incorporating a normal $2\pi$-periodic current component into our generalized Fokker-Planck analysis not only preserves the multiple critical currents we identified previously at high bath temperature, but actually leads to additional current regimes and hence finer structure in the interference spectra.  Figure \ref{ConventionalCurrents} illustrates the voltage $V$ and its second derivative $d^2 V/dI^2$ as a function of $I$ and $f$, along with specific voltage-current line cuts, for different conventional current strengths.  The following points made in the main text are worth reiterating: $(i)$ increasing $I_{2\pi}$ washes out the $\pi$-junction-like behavior in Fig. 2(d) of main text, $(ii)$ with `large' $I_{2\pi}$ the interference pattern resembles that at low bath temperature, and in particular features lifted nodes, and $(iii)$ most importantly multiple critical currents remain visible even for substantial conventional currents.

\subsection{3. Comparison with the experiment of Ref.\ \cite{HgTeProximity}}

Reference \cite{HgTeProximity} studied the critical current of an extended superconductor-HgTe-superconductor junction similar to the setup we analyzed in our paper.  The experiment measured the current as a function of the magnetic flux through the junction, for different gate voltages applied to the HgTe spacer region.  The gate voltage tunes the HgTe barrier from a metallic state into the QSH-insulator phase by depleting the bulk carriers---thus changing the bulk of the system from an SNS to SIS junction.  The QSH insulator supports conducting edge channels, however, so that current flow between the superconductors arises mainly through the edges. More precisely, as shown in the main text the current is expected to be mediated (at least in part) by a pair of hybridized Majorana fermions that yield fractional Josephson currents.  In the experiment, the transition between the metallic and QSH states of the HgTe is accompanied by a change in the interference pattern of the critical current as a function of magnetic flux piercing the junction.

The data analysis performed in Ref.\ \cite{HgTeProximity} clearly reveals the transition between a uniform current flowing through the junction in the metallic state, and edge currents dominating the flow in the QSH phase.  Notably, the critical current extracted from measurements in the QSH regime does not vanish at any value of the magnetic flux (but see remarks at the end of this section).   This observation is certainly intriguing given our prediction that similar behavior can emerge from topological superconductivity when parity switching occurs in the low-bath-temperature limit.  We caution though that drawing firm conclusions requires further experiments.  As emphasized in the main text node lifting can arise from other sources such as current asymmetry.  Indeed, this is how Hart \emph{et al}.\ interpreted the lifted nodes in their data.  Investigating how the nodes evolve as a function of an in-plane magnetic field  is one way of distinguishing our node-lifting mechanism from other, more conventional sources.

Next, we wish to emphasize the regime of validity of our theory and compare with experimental parameters from Ref.\ \cite{HgTeProximity}. Our predictions are expected to be relevant for extended QSH Josephson junctions satisfying the following properties:

$(i)$ The junction should be overdamped. In other words, the capacitance of the system, which is determined by the geometry of the junction, should be small. The precise condition is $\left(2e^2\Delta{R^2}C/\hbar^2\right)^{1/2}\ll1$, with $\Delta$ being the superconducting gap and $R$ the system's normal resistance.   In general, the opposite underdamped regime can be easily recognized in experiments, since it is characterized by hysteresis in the $I-V$ curves. In the experiment performed in Ref.\ \cite{HgTeProximity}, hysteretic  $I-V$ curves appear when the HgTe is metallic, but disappear as the system is tuned toward the transition into the QSH regime indicating the onset of overdamping.  Following our theory, it is possible to find the retrapping currents of a topological Josephson junction for different values of the magnetic flux.

$(ii)$ In order to maintain well-defined 1D topological superconductors across the barrier, the junction length $L$ (recall Fig.\ 1 from the main text) must significantly exceed the induced superconducting coherence length $\xi$.  If this is not the case, the Majorana modes in each side of the junction interact; the existence of the fractional Josephson currents that we invoked then becomes suspect. In the setup of Ref.\ \cite{HgTeProximity}, the length of the junction is $L \approx 4\mu$m. The superconducting leads are made of aluminum which has a characteristic coherence length of about  $1.5\mu$m in the bulk. Thus, although we do not know the induced coherence length, it is plausible that it is smaller than the junction length.

$(iii)$ The width $W$ of the junction should be much larger than the Fermi wavelength of the superconductors.  This condition is required to ensure that tunneling of electrons between the two superconductors via the bulk is negligible. Unlike the edges, the bulk of the QSH system is insulating, and current can flow through it only by two electron tunneling.  Bulk supercurrents may be even further suppressed by fabricating geometries where the junction width is smallest at the edges of the system (by, say, employing horseshoe-shaped superconductors). We wish to emphasize that although we assume in our calculation that the width of the junction is smaller that the superconducting coherence length, we expect similar physics to emerge even outside of the $W \ll \xi$ narrow-junction limit.   Larger junction widths yield a modified current-phase relation for the fractional Josephson currents as discussed recently by Beenakker \emph{et al}.\ \cite{BeenakkerWideJunction}.  These fractional Josephson currents should nevertheless still yield similar unconventional interference spectra characteristic of topological superconductivity, with or without parity switching.

Finally, it is worth noting that Ref.\ \cite{HgTeProximity} employed a voltage cutoff to determine the minimum critical current and used a conventional current-phase relation $\propto \sin\phi$ to extract the current profile from the interference pattern.  We stress that because topological superconductivity emerges very naturally in the quantum spin Hall system, it is not obvious how such a conventional current-phase relation can arise in the depleted regime.  It would thus be extremely interesting to revisit the current-voltage characteristics obtained in Ref.\ \cite{HgTeProximity} and attempt a quantitative fit assuming the (more natural) topological superconducting scenario.  To do so simulations of the full 2D system would be required to account for the finite width of the edge states, which would produce a decaying envelope of the critical current with magnetic flux as observed experimentally.  The numerically determined current-phase relation could then be inserted into the Fokker-Planck equation to incorporate the effects of parity switching.  If quantitative agreement can be obtained, this would by itself lend compelling support to the onset of topological superconductivity in the HgTe junction.

\bibliography{HgTeJJ_refs}

\end{document}